\documentstyle[aaspp4]{article}
\begin{document}
 
\lefthead{Elmegreen, Efremov, \& Larsen}
\righthead{A Young Globular Cluster in NGC 6946}
\slugcomment{in press {\it Astrophysical Journal}, Vol 535, June 1 2000}
 
\title{A Young Globular Cluster 
in the Galaxy NGC 6946\footnote{Based on
observations made with the Nordic Optical Telescope, operated on the
island of La Palma jointly by Denmark, Finland, Iceland, Norway, and
Sweden, in the Spanish Observatorio del Roque de los Muchachos of the
Instituto de Astrofisica de Canarias.}}

\author{ Bruce G. Elmegreen\altaffilmark{1},
Yuri N. Efremov\altaffilmark{2}, and
S{\o}ren Larsen\altaffilmark{3}}

\altaffiltext{1}{IBM Research Division, T.J. Watson Research Center,
P.O. Box 218, Yorktown Heights, NY 10598, USA, bge@watson.ibm.com}
\altaffiltext{2}{P.K. Sternberg Astronomical Institute, MSU,
Moscow 119899, Russia, efremov@sai.msu.su} 
\altaffiltext{3}{Copenhagen University Astronomical Observatory,
Juliane Maries Vej 32, 2100 Copenhagen {\O}, Denmark}

\begin{abstract} A globular cluster $\sim15$ My old that contains
$\sim5\times10^5$ M$_\odot$ of stars inside an $\sim11$ pc radius has
been found in the nearby spiral galaxy NGC 6946, surrounded by clouds of
dust and smaller young clusters inside a giant circular bubble 300 pc in
radius. At the edge of the bubble is an arc of regularly-spaced clusters
that could have been triggered during the bubble's formation. The region
is at the end of a spiral arm, suggesting an origin by the asymmetric
collapse of spiral arm gas. The globular is one of the nearest examples
of a cluster that is similar to the massive old globulars in the Milky
Way. We consider the energetics of the bubble and possible formation
mechanisms for the globular cluster, including the coalescence of
smaller clusters.

\end{abstract}

Subject headings: stars: formation ---
ISM: bubbles --- globular clusters: general 

submitted ... 7 June 1999, accepted ... 18 January 2000

\section{Introduction} During a recent search for massive young clusters
in the nearby spiral galaxy NGC 6946, Larsen \& Richtler (1999) found a
circular bubble containing numerous small clusters and a bright compact
cluster that resembled a young globular. Some of the small clusters in
the bubble are organized into arc-like shapes. This feature was found
and sketched first by Hodge (1967) as the only positive detection in his
search for multiple-arc structures similar to that in the Constellation
III region of the LMC. The diffuse object inside the bubble was noted
later by Efremov (1999), who suggested it was a cluster causally related
to the arcs in the same way that the LMC massive cluster NGC 1978 might
be related to Constellation III.

The mass and radius of the brightest cluster in this region make it
similar to the globular clusters in the outer halo of the Milky Way (see
review in VandenBerg, Stetson \& Bolte 1996). Globular cluster formation
is very rare in normal galaxy disks (the Milky Way disk has none), yet
other young globulars have been found in dwarf galaxies like the LMC
(van den Bergh 1991; Richtler 1993), and in interacting galaxies
(Whitmore \& Schweizer 1995) and starbursts (Maoz et al. 1996; Holtzman
et al. 1992; Meurer et al. 1995). Because of this, there is growing
anticipation that we will soon be able to understand the origin of the
old population of globular clusters in the Milky Way and other galaxies
by studying their younger versions. In this respect, we can use the
young globular in NGC 6946 for closer study.

Here we consider the properties of the bright compact cluster in NGC
6946 in view of the environment in which it formed. The observations are
presented in section 2, an analysis of the energetics and pressures of
the region are in section 3, and a discussion of possible mechanisms for
the formation of the globular are in section 4. 

\section{Observations}
\label{sect:obs}

Observations of clusters in NGC 6946 and other galaxies were made with
the Nordic Optical Telescope by Larsen \& Richtler (1999), who discussed
the results in general terms. The bubble surrounding the brightest
cluster in the galaxy caught our attention because of its peculiar
structure and possible implications for understanding globular cluster
formation. 

\begin{figure} 
\vspace{6.in}
\includegraphics{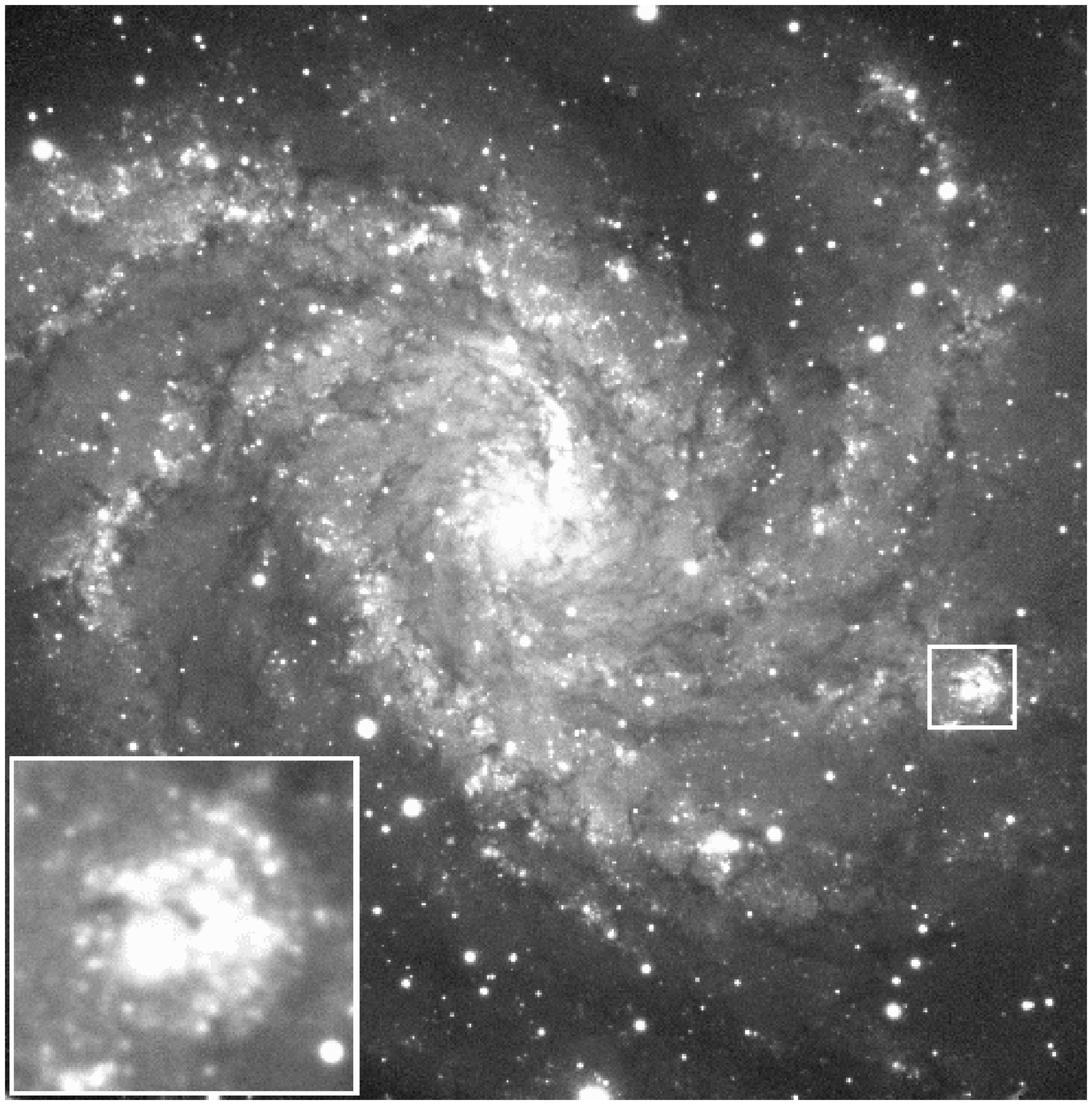}
\caption{A color image of NGC 6946 from the Nordic
Optical Telescope, with the bubble outlined, and an insert of the
bubble on a larger scale.  The bubble coordinates are in
Efremov (1999).  The globular cluster is the bright source just to the
lower left of center in the insert.  Numerous arcs of clusters can be
seen at the periphery and inside the bubble. The image is made from a
composite of B band for blue, V band for green, and R band plus
H$\alpha$ for red.  North is up. } \label{fig:color}\end{figure}

Figure \ref{fig:color} shows a color\footnote{Black and White for Astro-ph
to limit file size} 
image of this galaxy in B, V, and
R+H$\alpha$, with an insert magnifying the bubble region. The structure
of the star formation here is unlike anything else in the galaxy. It is
located at the end of a short spiral arm, downstream from a dust lane,
with a circular bubble outlined by an arc of clusters in the west and
a dark region, suggesting dust, just to the right. Red patchy emission
from H$\alpha$ surrounds the bubble as if star formation is continuing
there in dense residual gas. The brightest and largest concentration of
starlight emission inside the bubble is the globular-like cluster. The
overall dimensions of the cavity are $\sim600$ pc east-west and $\sim550$
pc north-south, assuming a distance of 5.5 Mpc (Tully 1988).

\begin{figure} 
\vspace{6.in}
\includegraphics{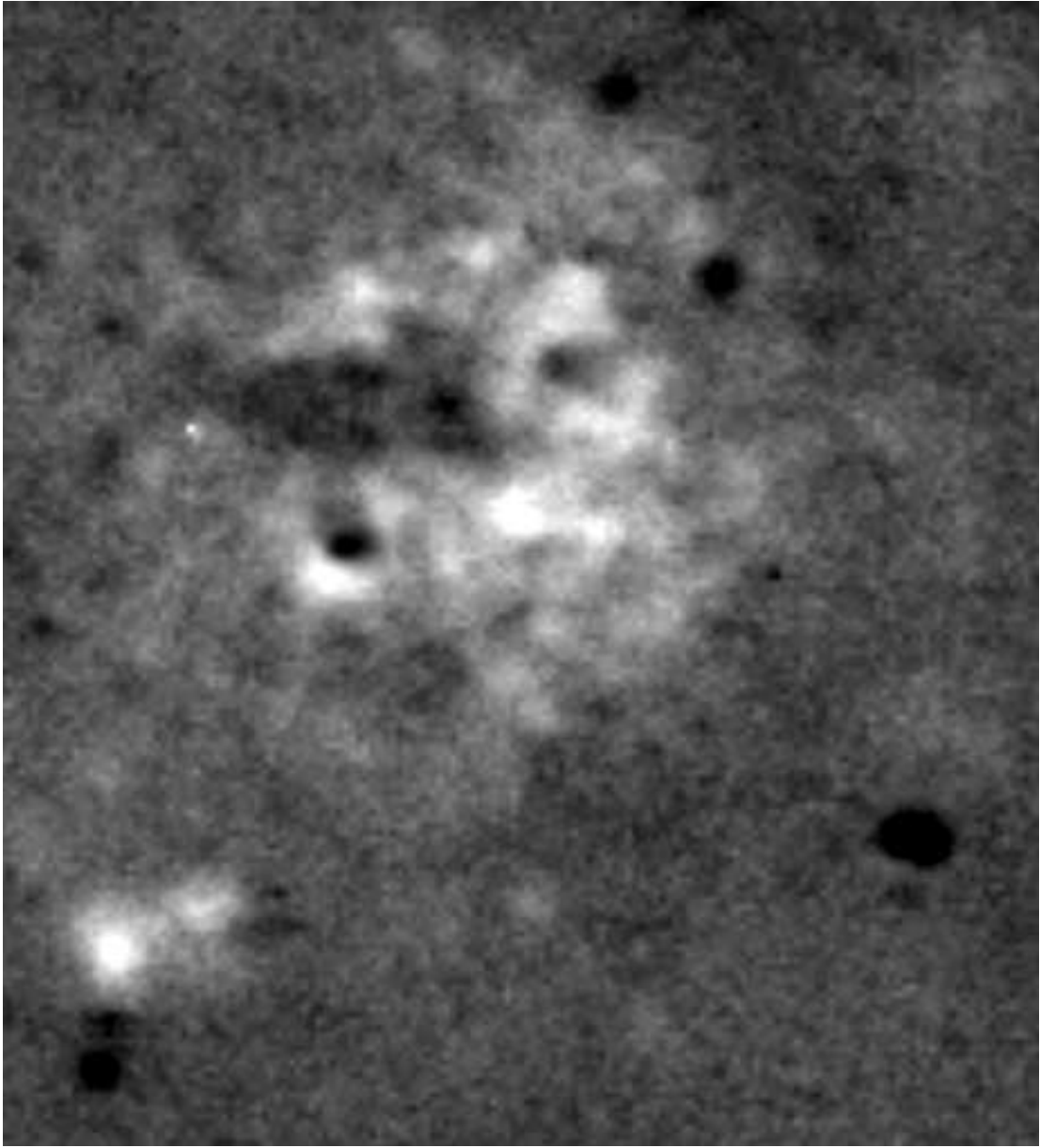}
\caption{A V-I image of the bubble region of
NGC 6946 showing probable dust features as dark regions.} \label{fig:vi}
\end{figure}

The extinction in the region is shown better by a plot of V$-$I in figure
\ref{fig:vi}, which covers 28''$\times$33'' around the bubble. Many of the
dark patches are irregularly shaped, like clouds in the midst of diffuse
and clustered stellar emission. A large dark region near the center
of the bubble, north of the brightest cluster, could be the remains of
the molecular cloud from which the cluster formed. Its dimensions are
$\sim275\times80$ pc. We measured the extinction in this region in two
ways: the brightness deficiency in the B band, and the V-I color excess,
both compared to the surrounding light. These two measurements are
independent and they give the same result, $A_B\sim1$ mag.. This implies
that the feature is really dust and not just a lack of stars, and that the
dust is mostly in front of the emission. With this average extinction for
the molecular cloud feature and its size given above, the 
mass becomes $\sim2\times10^5$ M$_\odot$.

Another V$-$I dust feature halfway out to the western edge resembles a
cometary globule, and this, along with the scalloped structures along
the western edge, point back to the brightest cluster. Other dust
patches and partial arcs are also evident in the V-I figure, inside the
bubble and beyond the bubble boundary. The dark region just to the north
of the globular cluster is probably an artifact of color variations in
the point spread profiles of the bright cluster image. 

Figure \ref{fig:extcorr} shows a V-band image of the same bubble region
on the left and a de-reddened V-band image on the right, made from
the observed ratios of V and I band intensities and corrected in V for
the corresponding extinction, assuming all the dust is foreground and
the background color is uniform. Many of the dark regions and bands in
figure \ref{fig:vi} fill in with what appear to be tiny clusters, and
some of the arc-like structures go away. This technique cannot correct
for extremely heavy obscuration because then there is no emission for the
determination of V-I color excess. The dark region beyond the bubble in
the west, seen in figure \ref{fig:vi}, may have this problem (the spots
indicate that the V/I intensity ratio has overflowed).

A low-contrast I-band image with cluster ages in millions of years is
shown in figure \ref{fig:ages}.  The ages were determined from their
integral colors, U-B and B-V, following the S-method and calibration in
Girardi et al. (1995). For the brightest clusters, the ages range from
10 to 30 Myr with no overall pattern inside the bubble. The colors of
the globular cluster (U-B=-0.74, B-V=0.01, V-I=0.52; Larsen 1999), give
it an age of $\sim15$ million years.  The bright spot in the southeast
outside the bubble is a foreground star.

A color-magnitude diagram for the same objects is shown in figure
\ref{fig:cmd}.  The lines are stellar isochrones for stars at the indicated
ages, with each line spanning a wide range of stellar masses from 0.6
M$_\odot$, which is outside the figure, to the masses corresponding
to the endpoints of stellar evolution (from Bertelli et al. 1994).
The diamonds are the objects inside the bubble; the source at $V=-12.9$
is the globular cluster.  This figure demonstrates that the small sources
inside the bubble are too bright to be stars if they are at the
distance of NGC 6946.  Even the brightest stars in
M33, at M$_V\sim-9$ mag (Humphreys \& Sandage 1980), are 
not as bright as the objects in the bubble.  

The ages given in figure \ref{fig:ages}
are slightly sensitive to unknown absorption. We
can estimate the error from figure 10 in Girardi et al. (1995), which
suggests that a color excess of E(B-V) = 0.2 mag. changes an age of 15
My to 11 My. Thus the ages determined from the S-method are probably
accurate to no better than $\pm30$\%, and may be systematically large. We
can also estimate the ages from the extinction-free $Q$ parameter ($Q =
(U-B) - 0.72 * (B-V)$; van den Bergh 1968), calibrating Q vs.  log(age)
using the population synthesis models by Bruzual \& Charlot (1996),
and assuming all the emission is behind the dust and not mixed with
it. These ages also tend to be slightly less than the values given in
figure \ref{fig:ages}.  For example, the globular cluster age becomes
$\sim 14$ My from its extinction-free $Q$ value. This second check on the
ages is also useful because the Girardi et al. ages were calibrated for
the LMC, where the metallicity is relatively low.  Considering measurement
errors and extinction uncertainties, the ages in figure \ref{fig:ages}
are probably accurate only to $\pm10$ My. In what follows, we assume
an age for the whole region of $\sim30$ My; it could range between 20
and 40 My, but this distinction is not important. The energy required
to make the bubble scales as the inverse square of the age.

The absolute magnitude ($V=-12.9$ mag) of the globular cluster makes
its mass $6.9\times10^5$ M$_\odot$, from the calibration in Bruzual \&
Charlot (1993, 1996; see Leitherer et al. 1996), assuming a Salpeter IMF
between 0.1 M$_\odot$ and 150 M$_\odot$. If we assume a more realistic
flattening of the IMF below $\sim0.3$ M$_\odot$ (Festin 1997;
Hillenbrand 1997; Luhman \& Rieke 1998; Lada, Lada \& Muench 1998;
Hillenbrand \& Carpenter 1999), then the mass is $\sim5\times10^5$
M$_\odot$. Other clusters in the same region are at least 2 magnitudes
fainter than the globular, although they have about the same ages, and all
the other clusters elsewhere in this galaxy are at least 1 magnitude fainter
than the globular.  

The radial intensity profile for the globular cluster is shown in
figure \ref{fig:profiles}, along with the profile for the bright star
in the southwest.  The globular is clearly more extended than the star.
The half-light radius is 11 pc from a fit to a King light profile,
corrected for telescope resolution (Larsen \& Richtler 1999; Larsen
1999). This radius is highly uncertain because the seeing resolution of
0.6'' corresponds to 15 pc.  The radial profiles of the fainter objects
in the field are somewhat extended too, making them appear fuzzy in the
image and more like clusters than stars. However, the S/N for these
profiles is too poor to give cluster sizes.  We intend to verify the
cluster nature of the fainter objects using HST images.

The large-scale environment of the bubble is shown in figure
\ref{fig:env}, from an I-band image. The bubble is clearly defined in the
midst of spiral arm stars and surrounding dust. There is a thin dust arc
just outside the bubble edge in the southwest, which is probably part of
the dust shell surrounding the cavity. A larger dust feature is at the
edge of the bubble in the northwest, connecting to the spiral arm dust
lane that trails off further in this direction (cf. Fig \ref{fig:color}).
Essentially all of the star formation in the vicinity of the globular
cluster is confined to the bubble cavity; the surrounding dust hides
only a small amount of additional star formation in comparison to
what has already occurred, considering the weak H$\alpha$ outside the
bubble (Fig. \ref{fig:color}).  This suggests that much of the star
formation inside the bubble, including the arc of clusters at the edge,
is causally related and possibly triggered or synchronized by a series
of high pressure events.

\section{The Origin of the Bubble}

\subsection{Multiple stellar winds and supernovae}

\begin{figure} 
\vspace{4.in}
\includegraphics{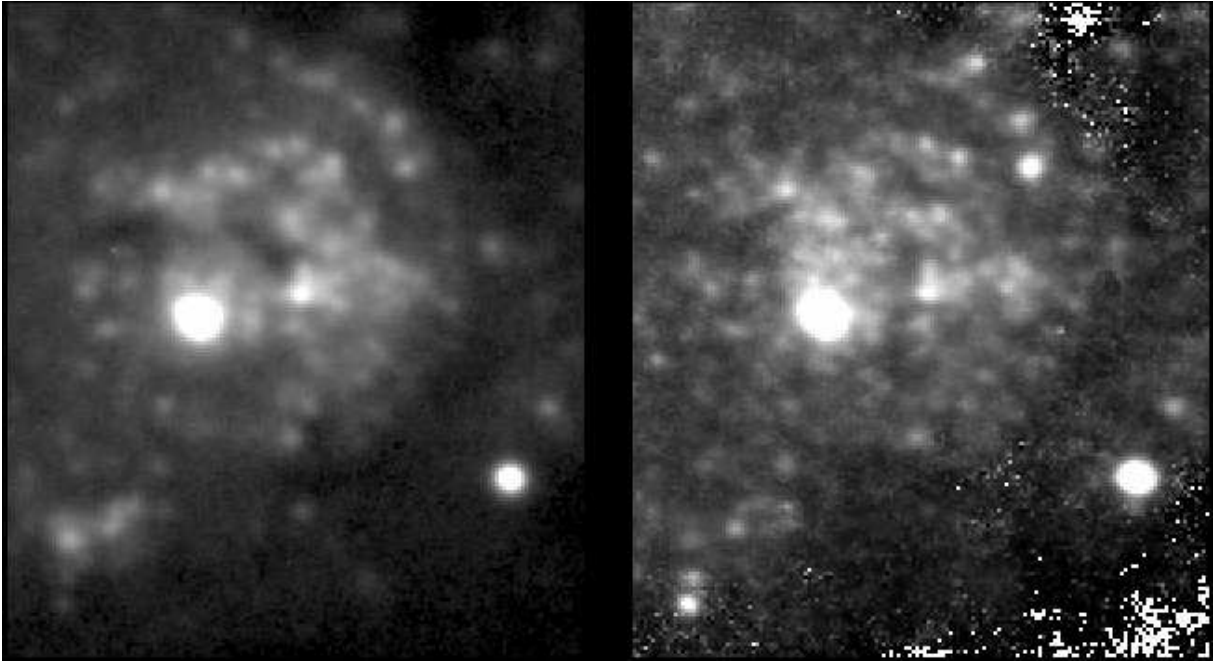}
\caption{Left: A V-band image of the
globular cluster region. Right: An extinction corrected V image
made from the V and I-band images, showing numerous emission sources 
in a dense core around the globular cluster. }\label{fig:extcorr}
\end{figure}

\begin{figure} 
\vspace{6.in}
\includegraphics{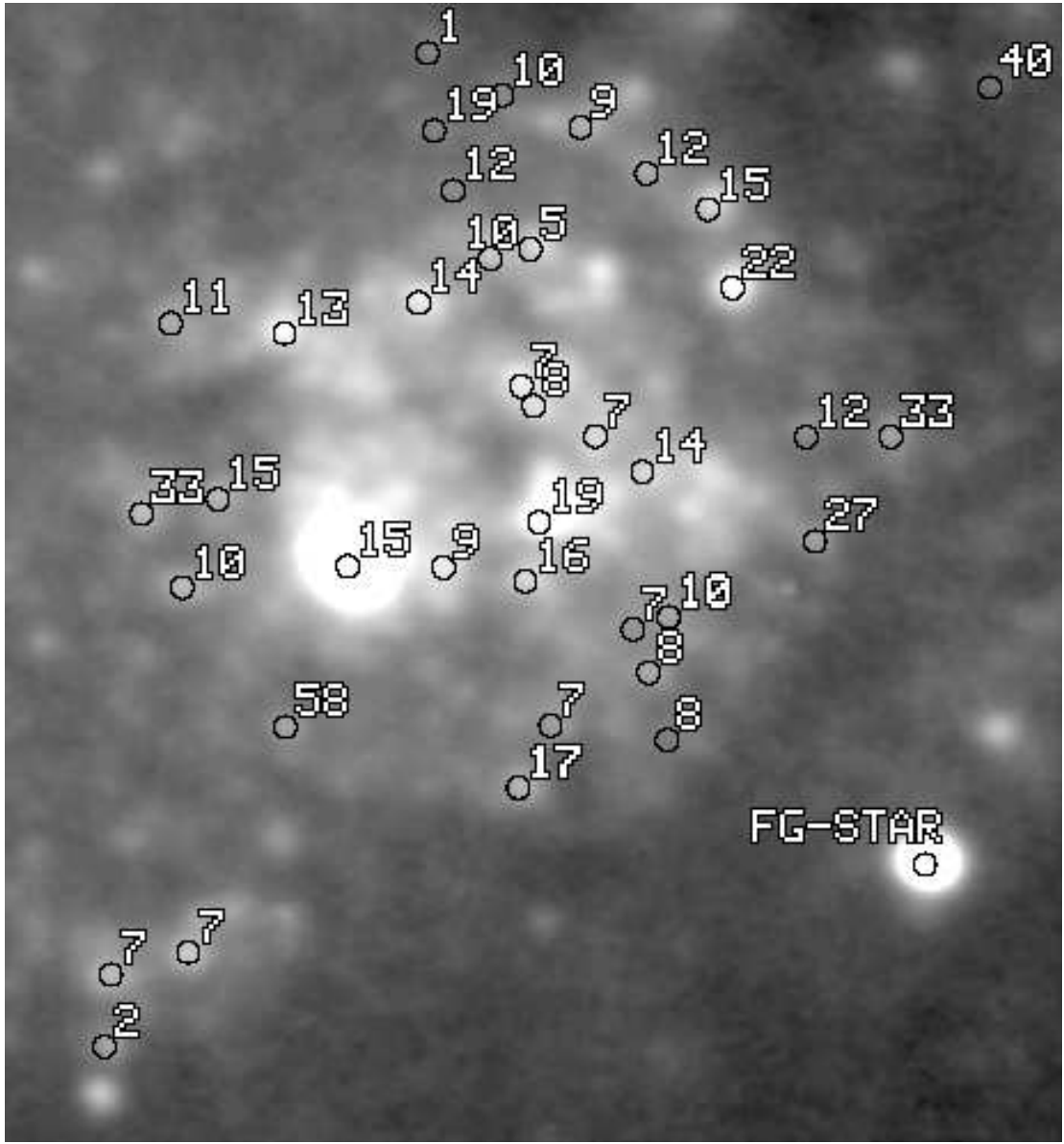}
\caption {A low contrast I-band image of the bubble
region with the cluster ages in Myr, calculated using the
method of Girardi et al. 1995. } \label{fig:ages}\end{figure}

\begin{figure} 
\vspace{4.in}
\includegraphics{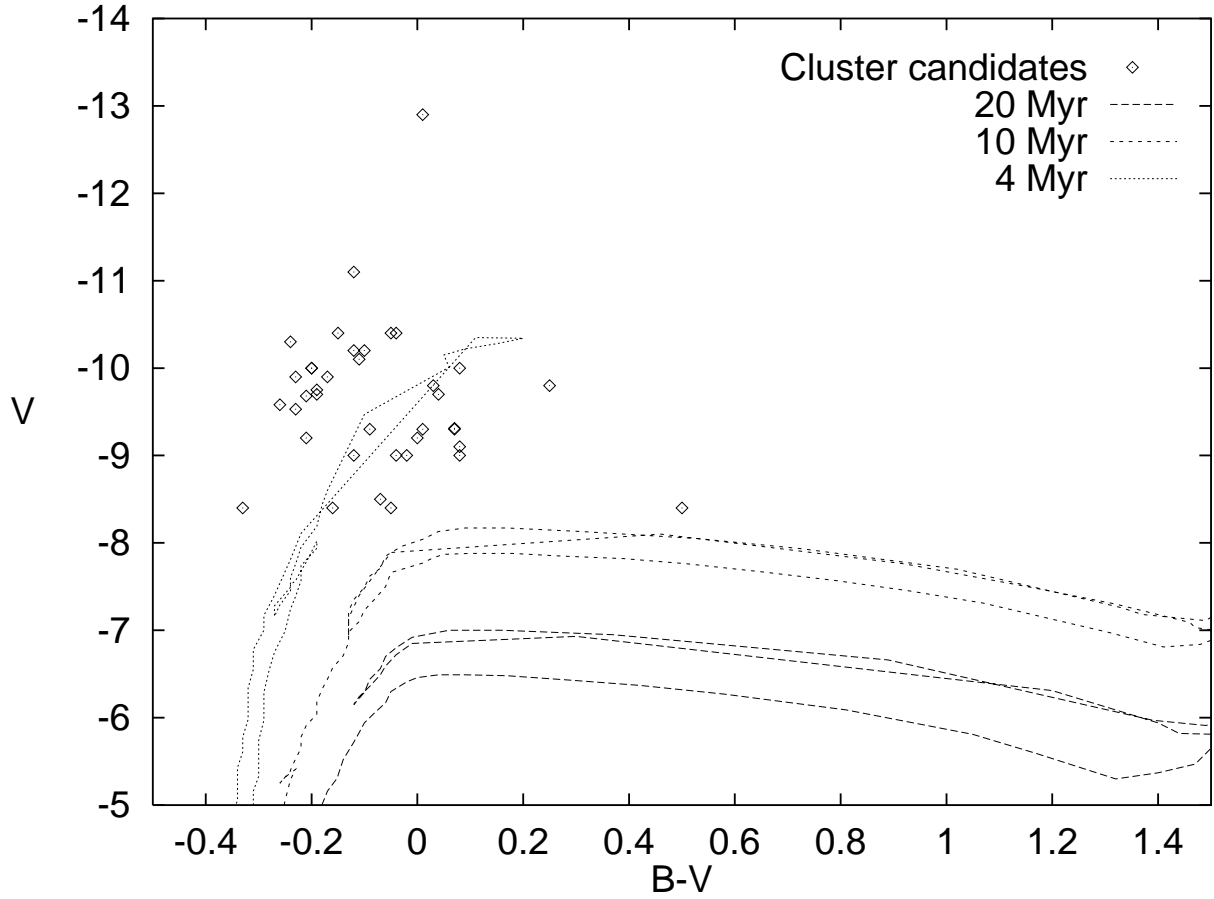}
\caption {A color-magnitude diagram for the 
clusters in the bubble region whose ages are shown in figure
4. The lines are isochrones with the indicated ages.
Most of the objects are brighter than the brightest stars in a
typical galaxy, which confirms that they are clusters.} 
\label{fig:cmd}\end{figure}

\begin{figure} 
\vspace{4.in}
\includegraphics{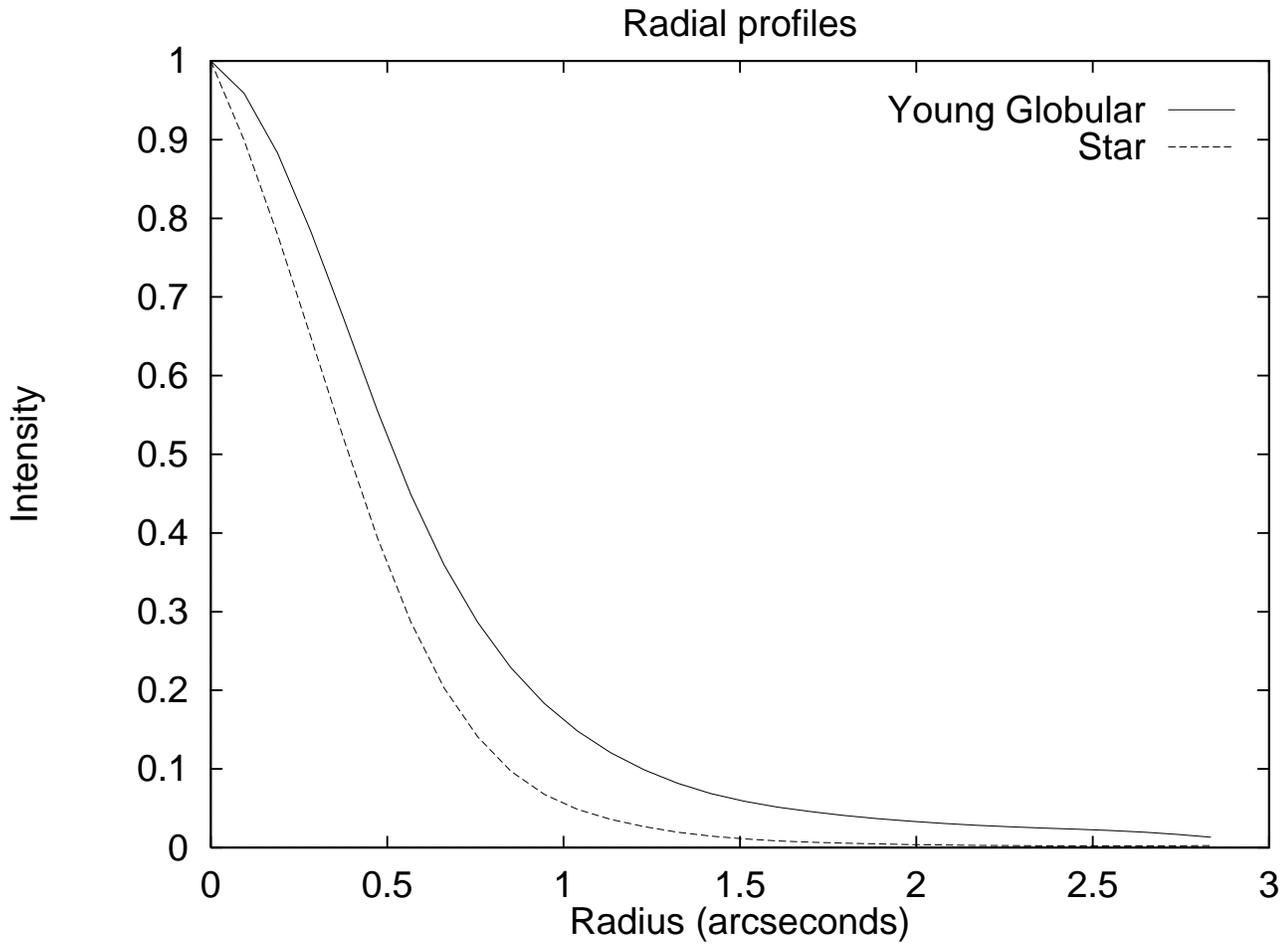}
\caption {Radial intensity profiles in V band for
the globular cluster and the bright star in the southwest. 
The fit to a globular cluster half-light radius of 11 pc
corrects for the image resolution of 0.6''.}
\label{fig:profiles}\end{figure}

\begin{figure} 
\vspace{6.in}
\includegraphics{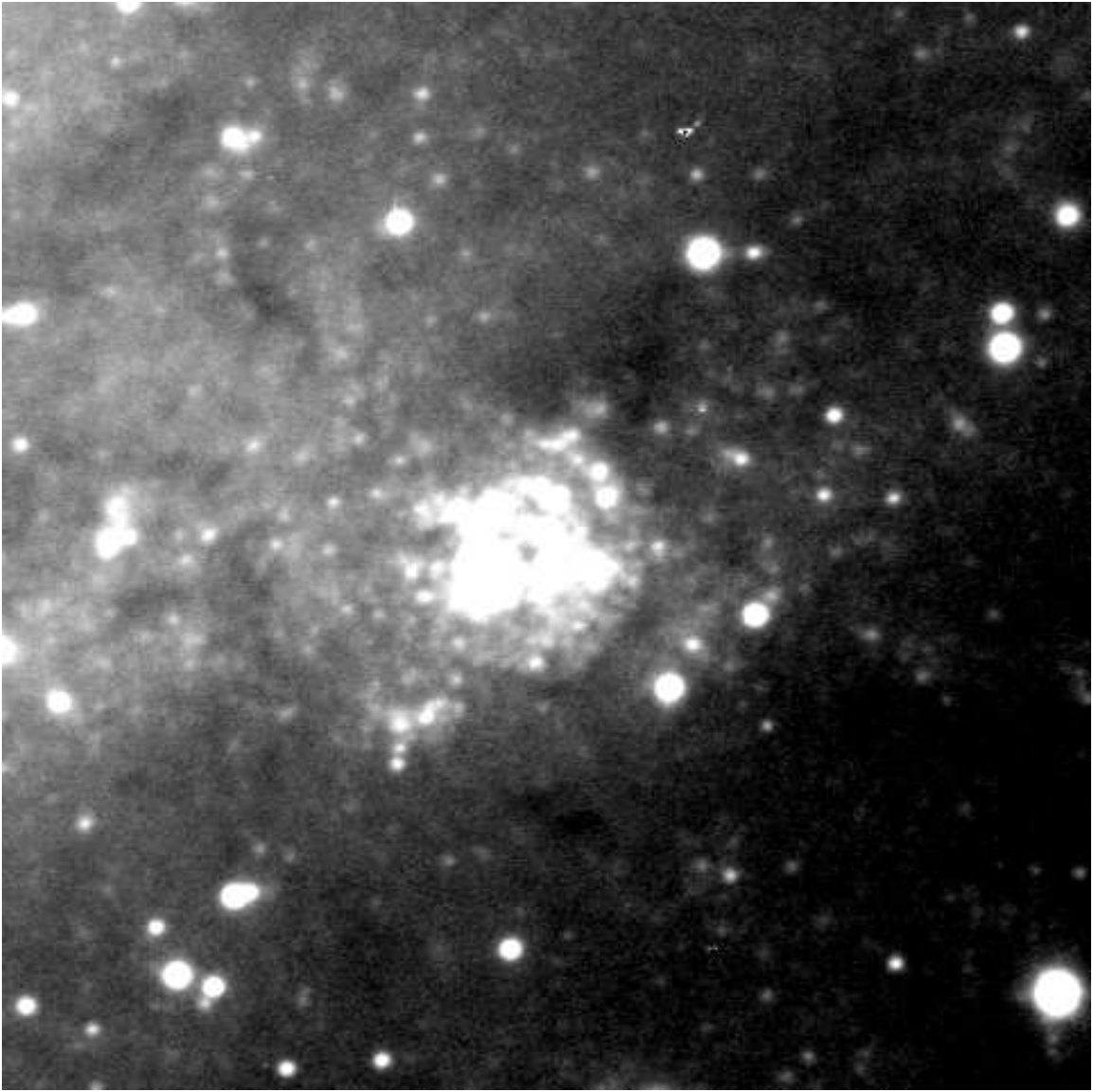}
\caption{An I-band image showing the field around
the bubble and globular cluster. The nearly perfect circular
shape and extreme concentration of star formation inside the
bubble are unusual.}  
\label{fig:env}
\end{figure}

Figures \ref{fig:vi}--\ref{fig:env} 
indicate that the bubble in NGC 6946 contains a single
bright cluster and a concentration of $\sim50$ smaller sources, many of
which are probably smaller clusters.

Most of these clusters are
concentrated within a $\sim300$ pc diameter region inside a $\sim600$ pc
cavity. A few other clusters lie along the rim of the bubble and more
may be outside, producing the H$\alpha$ emission evident in 
figure \ref{fig:color}.
The densest concentration of sources is within $\sim60$ pc of the
globular, as seen in figure \ref{fig:extcorr}. 
We count about 25 objects there. This concentration is packed
so tightly that many of the luminous sources blend together.

The total luminosity of all of the bright clusters in the bubble
suggests that a large number of supernovae exploded during its $\sim30$
My lifetime. These supernova and the winds of the associated massive
stars could have dispersed most of the dense gas that formed the
clusters and left only the $\sim10^5$ M$_\odot$ dust cloud and other
dust clouds seen in figure \ref{fig:vi}, now located $100$ pc or more from the
globular. The cometary and shell-like debris inside the cavity could
have been shaped by the supernova and stellar winds too.

A disturbance $R\sim300$ pc in radius that has been driven for $t\sim30$
million years by continuous pressures from winds and supernovae will
have a mass $M\sim\pi R^2\Sigma$ for ambient surface density $\Sigma$
and a speed $v=0.6R/t$ (Weaver et al. 1977). The kinetic energy is
$E=0.5Mv^2\sim 0.5 \pi R^2\Sigma \left(0.6R/t\right)^2=
2.1\times10^{51}A_V$ erg. Here we have rewritten the ambient mass column
density of the galaxy in terms of the equivalent V-band extinction
through the disk, $\Sigma=4.5\times10^{-3}$A$_V$ gm cm$^{-2}$, using the
conversion in Bohlin, Savage \& Drake (1978) with 10\% He by number.

Typically the energy required to make a bubble from multiple supernovae
and stellar winds is $\sim10$ times the kinetic energy of the expansion
because of energy loss from radiation in the shocks and decompression of
the cavity (MacLow \& McCray 1988). For example, the product of the wind
luminosity and the time in the Weaver et al. (1977) model (cf. their
equation 21) is 5.1 times the instantaneous kinetic energy. For a factor
of 10 from MacLow \& McCray, the source energy becomes $10^{52.3}A_V$
ergs, scaling with the inverse square of the age of the bubble.

We expect $A_V\sim3$ before star-formation began for the average of the
dense part of this spiral arm gas, based on direct measurements of
extinction in spiral arm shocks (Elmegreen 1980) and on extinction, HI,
and CO maps of this galaxy. The average extinction on $\sim100$ pc
scales surrounding the bubble currently exceeds 2 magnitudes in B band
(Trewhella 1998), while the summed atomic (Boulanger \& Viallefond 1992)
and molecular (Tacconi \& Young 1989) gas column densities over a 1 kpc
region correspond to $A_V=1.1+1.4=2.5$ mag.

If $A_V$ equaled 3 mag in the vicinity of the bubble before
star-formation began, then the expansion required the equivalent of
$\sim60$ supernovae, each with a typical energy of $10^{51}$ ergs. If
the bubble age is only 15 My, the age of the globular cluster, then the
equivalent number of supernova would be $\sim240$. This is easy for a
cluster with $\sim5\times10^5$ M$_\odot$, because there would be 2700
stars more massive than the turnoff mass at 15 My (which is 14
M$_\odot$), assuming a flattened Salpeter IMF out to 150 $M_\odot$ and
stellar evolution models in Bertelli et al. (1994). With so many
expected supernovae, the expansion age could be even younger than 15 My,
or $A_V$ larger than 3. Presumably some of these supernovae made the
smaller shells inside the bubble too, producing the overall frothy
appearance in figure \ref{fig:color}.
In this case these supernovae may have been ejected from the globular.

The average energy density in the bubble is $\sim10^{52.3}A_V/\left(4\pi
R^3/3\right)\sim 10^{4.6}A_V$ cm$^{-3}$ K. The energy density in the core,
at the peak of the input rate, should have been at least 10 times this
average. For $A_V\sim3$, this peak is $1.2\times10^6$ cm$^{-3}$ K, which
is similar to that around OB associations in the Solar neighborhood.
Thus the pressure that made the bubble was not unreasonably large, in
spite of the high density of stars in the center. This is because the
overall star formation rate in the $\sim600$ pc region inside the bubble
is typical for spiral arms. The only difference in the bubble here is
that most of this star formation went into a single dense cluster,
rather than a giant, extended star complex, like Gould's Belt.

\subsection{Alternative Scenarios}

An alternative source of pressure for the bubble should also be
considered for this massive central cluster. It seems possible that such
a cluster would have sufficient opportunity to produce several rare
hypernovae (Paczy\'nski 1998) from the collapse of extremely massive or
otherwise rare single or binary stars. For example, a Salpeter IMF that
flattens below 0.3 M$_\odot$ in a cluster that contains $5\times10^5$
$M_\odot$ would be likely to produce $\sim8$ stars with masses between
100 M$_\odot$ and 150 M$_\odot$ if the maximum mass were 150 M$_\odot$.
Such massive stars could produce hypernovae of the type described by
Paczy\'nski (1998). If the IMF continued up to arbitrarily high masses,
then the largest star would have an unlikely mass of $\sim900$ M$_\odot$
in a $5\times10^5$ M$_\odot$ cluster, just from sampling statistics
alone (without any consideration of whether such massive stars are
physically possible). Of course, stars this massive have never been
found, but neither has a fixed upper limit to the stellar mass in a
cluster: the largest star just keeps getting larger for larger clusters
(e.g., Massey \& Hunter 1998). Thus the largest star could conceivably
contain several hundred M$_\odot$. If such a star, or any other rare
object exploded, possibly making a gamma ray burst, it could have an
energy larger than a single supernovae by a factor of $\sim100$ (Rees \&
M\'ez\'aros 1998) and be able to produce one of the large arcs of gas
and dust individually (see also Efremov 2000).

Although there is no direct evidence for unusual pressure sources like
hypernovae, there are several remarkable properties of the bubble that
are not usually part of the standard, multiple supernovae/stellar wind
scenario (Tenorio-Tagle \& Bodenheimer 1988).

First, the bubble is not elliptical or elongated along the major axis of
the galaxy. This implies that it is not a circular ring in the midplane
(unless an intrinsic ellipticity fortuitously cancels the projection
effect), but more like a sphere with some thickness perpendicular to the
plane. Yet its diameter (600 pc) is much larger than the likely disk
thickness. This leads us to wonder how such a circular bubble could have
formed in the first place. Its uniform shape would seem to require a
very uniform external medium for the pressure to push around, yet
disturbances perpendicular to the plane generally propagate into
decreasing densities. Perhaps the shocked gas in the density wave really
does have such a large thickness, and what we were calling a GMC in
Section \ref{sect:obs} is really the near side of the bubble, far 
above the plane. 

Second, the clusters that lie along the bubble rim to the north, west,
and south are somewhat uniformly spaced, as if they were triggered by
the regular gravitational collapse of a swept-up shell (e.g. McCray \&
Kafatos 1987). However, the ages of these rim clusters are about the
same as the ages of all the other clusters inside the bubble, including
the globular cluster, so there is no evident time sequence or source of
pressure for this triggering. 
Efremov (2000) suggested that there had to
be some invisible older generation ($\sim50$ My) of clusters or
super-explosions that made all of these features at about the same time,
i.e., the overall bubble, the outer cluster arcs, and possibly the globular
cluster.  Alternatively, the photometric ages could be wrong. 

Third, the interior of the bubble contains dust (and presumably gas)
clouds, including a comet-shaped cloud half-way out to the western edge.
How can the bubble {\it outside} these clouds be so circular? Wouldn't
the comet and other interior clouds block the expansion along their
paths if the pressure source was near the bubble center? Does this imply
that the bubble formed before the comet cloud and that there were two
epochs of high pressure in this region? A similar situation seems to
occur in the Constellation III region of the LMC, where the HI cavity is
much larger than a triggered arc of stars (Efremov \& Elmegreen 1998).
The solution here might be simply that the dark debris in the interior
of the bubble is far above the plane on the near side, and not on the
line of sight between the central pressure source and the western edge. 

Fourth, there is apparently some star formation beyond the bubble, in
the H$\alpha$-emitting regions. This makes us wonder if the western arc
of clusters is not really an arc of stars but an arc of clear-viewing,
seen through a dust screen made up of dense material with its own arc
shape, just inside the stellar arc, and other dense material with an
arc-shaped inner border, just beyond the western edge 
(cf. Fig. \ref{fig:vi}). 
What is
the source of the inner gas shell in this interpretation? Can multiple
supernovae and stellar winds make such a shell inside a pre-existing
bubble cavity, or does it require a more concentrated giant explosion
as in the hypernova scenario discussed above? Did this additional
explosion also occur after the bubble formed, as suggested
above for the comet cloud?  

All in all,
the high concentration of star formation in this region, forming
what appears to be a dense {\it cluster of clusters,} is unusual for
non-interacting galaxies. The presence of background stars to the west
of the bubble in figure \ref{fig:env} 
suggests that the obscuration there is not so
large that it could hide a significant amount of additional star
formation, at least not comparable to what is in the bubble itself. This
means that the western edge of the bubble is only partly the result of
an end to the clearing of the cavity; it is also from a lack of bright
clusters beyond. If there was any triggering of star formation,
especially to make the arcs along the bubble edge, then this triggering
has stopped near the current limit of the bubble. The sharp western edge
of the bubble may then be explained if this is a triggered shell of
clusters.

The sharp edge and the perfect circular shape of the western rim of
the bubble is the property shared also by a number of other stellar arcs
in a few galaxies, first of all in the Constellation III region of the LMC.
The circular and sharp outer edges indicate triggered star formation
in the partial spherical shells, the occurrence of the multiple arcs near each
other being connected with the common origin of the pressure sources.
Most probably they were ejected from a  rich nearby cluster (Efremov, 2000).
As mentioned above, the exact nature of suggested triggering is
difficult to understand, as is the source of pressure, but
super-explosions cannot be overlooked.

Thus we see that even though the energy and pressure inside the bubble
are somewhat normal for an aggregate of stars like this, the morphology
of the whole region is baffling. Perhaps higher resolution and longer
wavelength observations will help clarify these questions.

\section{Speculations about Globular Cluster Formation}

\subsection{General Discussion of the Observed 
Environment for Globular Cluster Formation}

Why should a globular cluster form in this region of NGC 6946 and no
where else? This is a special location at the abrupt end or break-point
in a secondary spiral arm. When star formation began, there was a gross
asymmetry in the self-gravitational force of the spiral arm gas, with
most of the acceleration pointing inwards along the arm. Such asymmetry
would cause catastrophic collapse in a large mass of gas, possibly like
the end of a tidal tail (Barnes \& Hernquist 1992; Elmegreen et al.
1993). The total atomic (Boulanger \& Viallefond, 1992) and molecular
(Tacconi \& Young 1989) mass observed in a $\sim$kpc region around the
globular cluster is $\sim10^8$ M$_\odot$, which is $\sim10$ times higher
than the typical cloud mass from normal density wave triggering
(Elmegreen 1994). 

Not all disk globular clusters form at the ends of spiral arms, however,
and the other arm endpoints in NGC 6946 contain no globular clusters,
even though they often contain massive star-forming regions (cf. Fig.
\ref{fig:color}).  Something peculiar happened here to make the dense,
massive cloud that was necessary for the globule to have formed.

What is unusual about this region? The density of the globular cluster
alone is not particularly high. The average half-light density is only
$\sim200$ stars pc$^{-3}$, which is much smaller than the $\sim10^4$
stars pc$^{-3}$ in the Orion Trapezium cluster (Prosser et al. 1994;
McCaughrean \& Stauffer 1994) and Mon R2 (Carpenter et al. 1997).
The corresponding $H_2$ density would have been $\sim6\times10^3$
cm$^{-3}$ with a 25\% star-formation efficiency (e.g., Elmegreen et al.
2000), and this is similar to that in a GMC core. The main peculiarity
with the globular cluster is the large mass at this density. For an
efficiency of 25\%, the gas mass to make a $\sim5\times10^5$ M$_\odot$
cluster would have been $\sim2\times10^6$ M$_\odot$, which is comparable
to that of the largest GMCs in the Milky Way but here concentrated inside
several tens of parsecs. Such a high mass concentration means that the
virial velocity and pressure of the cluster-forming core were large.
A $M=2\times10^6$ M$_\odot$ cloud with a half-light radius of $R\sim11$
pc has a virial speed $\left(0.2GM/R\right)^{1/2}\sim12$ km s$^{-1}$
and a pressure $P\sim0.1GM^2/R^4$, which converts to $P/k_B=6\times10^8$
K cm$^{-3}$ for Boltzmann constant $k_B$.

The pressure would have been slightly lower in the surrounding gas.
If we consider the core of luminous sources surrounding the globular
cluster in figure \ref{fig:extcorr}, which has a $\sim60$ pc radius,
and estimate that the total mass in that core increased linearly with
radius as for an $r^{-2}$ isothermal density profile, then the mass there
would be $\sim5$ times larger than the globular core mass, or $\sim10^7$
M$_\odot$.  The average pressure would be $\sim2\times10^7$ K
cm$^{-3}$ and the virial speed still $\sim12$ km s$^{-1}$.

These virial speeds are comparable to the thermal speed of an HII
region, suggesting that the O-type stars which formed in the clusters
would have had some difficulty destroying the cloud cores. The core
pressures are also large enough to avoid complete ionization of the gas.
These are welcome checks on the most basic theory of globular cluster
formation: the cloud-binding energy densities and escape speeds have to
be large enough to keep the gas around to continue forming stars, in
spite of all the O star ionization, until the critical efficiency to
form a bound cluster is reached (Elmegreen \& Efremov 1997). 

The pressure in the 60 pc core region was $\sim10$ times higher than the
typical pressure in a local OB association and molecular cloud core and
$10^3$ times higher than the ambient pressure in the Solar neighborhood.
The pressure in the 11 pc core was $\sim100$ times higher than in a
local OB association/GMC core. Most likely the ambient pressure in the
NGC 6946 region was $\sim10$ times the ambient pressure in the Solar
neighborhood, and the extra factor of $100-1000$ for the cloud cores
over the ambient value was from the self-gravity of the material
surrounding it. Such overpressures are typical for star-forming clouds in
the Solar neighborhood. What might have given the
pre-cluster interstellar medium a pressure of $\sim2\times10^5$
cm$^{-3}$K over a region perhaps a kiloparsec in size? 

Considering the location of this region of star formation inside NGC
6946, we should probably concentrate on the effects of a spiral density
wave shock. For an incoming average HI density of 1 cm$^{-3}$, a shock
speed of $\sim35$ km s$^{-1}$ would have reached the required
$2\times10^5$ K cm$^{-3}$ pressure over a large region. Such a shock
speed is not unusual for a galaxy like this, and when combined with an
asymmetric collapse from the arm end, could have conspired to make the
pressure in a massive virialized cloud somewhat high. 

A second constraint on the formation of the globular cluster is its
small photometric age. Obviously it had to form very quickly, probably
within only several million years, to have such a young age with no
obvious dust or irregular structure remaining inside of it. The only
dense dusty material that we see nearby is a $\sim10^5$ M$_\odot$ cloud
some 100 pc away, which could be the remains of the cloud core that
formed it. Such a quick formation time follows from the most basic model
(e.g., Fig. 4 in Elmegreen \& Efremov 1997) if clouds typically convert
their gas into stars in only several dynamical time scales of the cloud
core. At the pre-cluster density in the globular core, the dynamical
time was $\left(G\rho\right)^{-1/2}\sim0.8$ million years. This is fast
enough to form the globular and clear away most of the debris within its
$\sim15$ My lifetime. Such a short time scale also implies that
supernovae in the cluster did not enrich the gas much before it was
cleared away. This helps us understand how the oldest globulars in the
Milky Way halo can have very low metallicities even though they must
have had massive stars.

\subsection{A Coalescence Model for the Formation of the Globular Cluster}

These aspects of the environment of the globular cluster in NGC 6946 are
all consistent with what we believe to be necessary for the formation of
such a massive object with a normal IMF, and they are also consistent
with the location of this cluster inside the galaxy. They do not
indicate how the globular actually formed, however. It need not have
formed by the monotonic collapse of a single cloud core, for example, as
in the standard scenario for cluster formation. Instead, it could have
formed by the agglomeration and accretion of smaller clusters in the
vicinity, provided the timescale for such accretion was short enough.
Indeed, there are several peculiarities of star formation here that
suggest such accretion might have actually happened.

First, there is an extremely high concentration of small luminous sources
that may be individual clusters surrounding the globular cluster. This is
shown best by the extinction-corrected image in figure \ref{fig:extcorr}.
We count $\sim25$ small objects within a radius of 60 pc, and estimate
that there are probably an equal number in the foreground and background
of the bright globular cluster image. This would make the current cluster
density $n_{cl}\sim6\times10^{-5}$ pc$^{-3}$ if all these objects are
small clusters. 

Second, the mass function of all the known clusters in this region is
discontinuous.  The globular cluster is $\sim10$ times more massive than
any of the other clusters. This is unusual for disk cluster systems,
which tend to have a continuous power-law mass distribution (van den
Bergh \& Lafontaine 1984; see the review in Elmegreen et al. 2000).
Discontinuous mass functions, with a single largest object having
a mass much larger than any other object, can arise from a runaway
accretion process as the gravitational cross section of the largest
object increases with mass.  This could have been important for old
globular clusters because their mass distribution differs from that of
disk clusters today (van den Bergh 1995).  In fact the old clusters have
a characteristic mass similar to that of the globular cluster in NGC
6946, so formation processes like agglomeration, which break away from
the power law distributions of normal disk clusters, might be important.

Third, the photometric age of the globular cluster, $\sim15$ My, is
about the average age of all the other clusters in the bubble, which is
14.4 My not counting the globular cluster itself, or 12.5 My if we omit
the 58 My-old cluster in figure \ref{fig:ages}.  This equality is to
be expected in a coalescence model: the photometric age of the globular
would not be its true age from some single star formation event, but the
luminosity-weighted average age of all of the clusters it accreted. The
true age could be the maximum age of the neighbors, because this marks
the beginning of all star formation in the region and the time when
the accretion process might have begun. This third point is uncertain,
however, because of the statistical errors in the age measurements.

Finally, the globular cluster has a low average density, lower than the
densities of embedded clusters that typically form near the Sun.
Accretion can do this because it puffs up a cluster with the kinetic
energy of the accreting pieces, giving it a lower average density than
any single piece. We discuss this effect in more detail below.

We first check the accretion model by calculating the accretion
rate for the small objects that are currently in the dense
region surrounding the globular. This rate is $n_{cl}v\sigma$
for surrounding cluster density $n_{cl}$ estimated above, cluster
velocity dispersion $v$, and capture cross section $\sigma$. The
cluster velocity dispersion is taken to be 12 km s$^{-1}$ from the
virial dispersion in the cluster core, derived above. The capture cross
section is \begin{equation} \sigma\sim\pi\left(R_{GC}+R_{cl}\right)^2
\left(1+{{2GM_{GC}}\over{R_{GC}v^2}}\right) \end{equation} for core
radii of the globular and incident clusters, $R_{GC}\sim11$ pc and
$R_{cl}$, globular cluster mass, $M_{GC}\sim5\times10^5$ M$_\odot$,
and gravitational focusing factor ($\approx3.7$) in large parentheses
above. For $R_{cl}<<R_{GC}$, $\sigma= 1400 $ pc$^2$. Then the accretion
rate is $n_{cl}v\sigma\sim1$ field cluster per My. If the true age
of the globular cluster is not the average photometric age of all the
accreted clusters but the total age range in the whole region, which is
twice this average value, then the globular had $\sim30$ My to accrete
smaller clusters from the dense swarm around it. Moreover, the initial
field cluster density was probably higher before the accretion began,
so the accretion rate could have been higher then too. Thus the globular
could conceivably have grown by accretion over a $\sim30$ My period from
an initial size comparable to that of one of smaller clusters in the area.

The potential errors involved in this calculation should be reiterated
because the observations are not good enough yet to be certain about
the coalescence rate. First, the globular cluster radius is close to the
seeing limit so the King-profile fit used to determine $R_{GC}\sim11$ pc
is inaccurate. The cluster could be smaller. Second, the small sources
surrounding the globular in figure \ref{fig:extcorr} could be image
noise, or, if not, then some of the fainter ones could be luminous stars
instead of clusters.  Third, the ages of the clusters are not well known,
so we cannot be certain that the photometric age of the globular is the
average of all the others in the neighborhood. Fourth, the gas mass that
formed the globular is not observed directly, but only inferred from
the need for cluster self-binding, which implies a high star formation
efficiency.  The first three of these uncertainties could be clarified
by higher resolution observations. There is no direct way to know what
the previous cloud was like.

Nevertheless, the possibility that the globular cluster formed by the
accretion of smaller clusters is intriguing, and, if true, could change
our way of thinking about globular cluster formation. Thus we seek another
piece of evidence from the low density of the globular cluster today.
Such a density might be expected in a coalescence model because of
the conversion of neighbor cluster orbital energy into globular cluster
binding energy.

Consider two clusters with masses $M_1$ and $M_2$ and gravitational
binding energies $-\alpha GM_1^2/R_1$ and $-\alpha GM_2^2/R_2$ for
constant $\alpha$ depending on the mass distributions. If these clusters
coalesce after an initial release at infinite distance and zero
velocity, then, because of energy conservation, the final cluster will
have a total energy $-0.5\alpha G\left(M_1^2/R_1+M_2^2/R_2\right)$.
After relaxation, this will equal the total energy of the new virialized
cluster, $-0.5\alpha G\left(M_1+M_2\right)^2/R_3$ for new radius $R_3$.
For initial cluster densities $n_1$ and $n_2$, the final cluster
density, $n_3$, becomes \begin{equation}
n_3={{\left(n_1^{1/3}M_1^{5/3}+n_2^{1/3}M_2^{5/3}\right)^3}\over
{\left(M_1+M_2\right)^5}}. \end{equation} If we consider that the
globular cluster formed by the successive accretion of $N$ identical
smaller clusters, each of density $n_0$, and write the ratio of the
current globular density to this initial density, $X=n_3/n_0$, then $X$
decreases iteratively with $N$ as \begin{equation}
X_{N}={{\left(1+X_{N-1}^{1/3}N^{5/3}\right)^3}\over{\left(1+N\right)^5}}.
\end{equation} Starting with $X_{N=0}=1$, this gives a sequence of
$X_N=1/(N+1)^2=1,$ 0.25, 0.111, and so on. Thus after the successive
addition of $N=9$ small clusters, a large cluster, 10 times as massive,
forms with $\sim1$\% of the initial density. This factor of $\sim10$ for
mass is about the ratio of the globular cluster luminosity to the
luminosities of the bright neighboring clusters (2.5 mag difference in
brightness), and $\sim1$\% is about the ratio of the globular cluster
stellar density to the density in a trapezium-type cluster, which is
presumably similar to the cluster densities in the NGC 6946
neighborhood. 

These simple checks suggest that the globular cluster in NGC 6946 could
have had enough time, $\sim30$ My, and currently has about the right
mass and density, for it to have formed by the coalescence of many of
the smaller objects, perhaps normal clusters, that seem to be nearby. The
photometric age of the globular would then be the average age of these
neighbor clusters, $\sim15$ My, as observed within the errors. During
this same total time of 30 My, the pressures from thousands of stellar
winds and hundreds of supernovae in the dense star-forming region made
the 600 pc bubble and pushed the residual dense gas to $\sim100$ pc
distance inside the bubble, leaving a $\sim2\times10^5$ M$_\odot$ cloud
and several cometary-shaped clouds.

\section{Conclusions}

A young globular cluster has been found inside a bubble of gas filled
with numerous smaller clusters at the end of a short spiral arm in the
nearby galaxy NGC 6946. The photometric age of the globular cluster is
probably between 5 My and 30 My; the best fit is $\sim15$ My. The mass
of the globular is $\sim5\times10^5$ M$_\odot$, and the half-light
radius was
estimated to be $\sim11$ pc by a fit to King model. This radius is less
than the seeing limit (16 pc) so the actual radius could be smaller. The
mass and size of the whole star-forming region correspond to such a high
velocity dispersion and pressure in the previous cloud core that
disruption by OB stars and stellar winds would have been difficult
before the supernova era began. This observation satisfies the most
fundamental requirement of bound cluster formation, that the star
formation process be able to continue unimpeded until a high efficiency
is reached. The total time for the cluster to form was probably very
short, considering the $\sim1$ My dynamical time in the dense part of
the cloud core. This time is consistent with the young age and relaxed
structure of the globular cluster today. 

The bubble surrounding the globular cluster could have been made by the
wind and supernova pressures from evolved massive stars in the cluster.
No additional pressure sources are needed, like the ``super-supernova''
postulated by Hodge (1967). Nevertheless, such unusual explosions might
be expected for such a massive cluster, and they may have contributed to
some of the arc-like structures in the gas and dust. Some of these
cluster arcs seem to be real, and then they could be triggered star
formation. The average pressure in the bubble is typical for the
pressures in OB associations near the Sun, and the bubble size is not
unusual for giant disturbances in galaxies. The near-perfect circular
shape of the bubble is unusual, though, considering its size in
comparison to the likely scale height of the disk.

This circular shape and the absence of clusters outside the western
parts of the rim of the 'bubble' suggest another possible interpretation:
that this is in fact not a cavity in absorption (a bubble), but
a spherical shell of triggered star clusters, like some other
arc-shaped regions of star formation in other galaxies (Efremov 2000).
This would remove the difficulties with the clearing model.
At any event, such a spherical feature is unusual,
and more so if the similar ages for all of the clusters inside the shell
are confirmed. Then Hodge's (1967) suggestion that there was 
a super-explosion
in this region will be more tenable.
The only alternative might be an infall of a group of clouds.

Also unusual is the formation of a Gould's Belt worth of star mass
inside a half-light radius of only 11 pc or less. We suggested that high
pressures and an asymmetric collapse from the associated spiral density
wave might have produced the right conditions for such concentrated star
formation, but why this region should differ from so many other sites of
star formation at spiral arm ends is unknown. Perhaps the pressure
source that formed the outer arc of clusters, if that came first, was
also involved with the formation of the globular cluster. 

The possible formation of the globular cluster by the coalescence of
smaller clusters was also discussed. For the assumed parameters, the
cluster accretion rate is high enough to explain the globular cluster
mass, and the gradual decrease in virial density by the addition of
moving sub-clusters can explain the current low globular cluster
density. Also explained is the apparent globular cluster age, which
seems to equal the average age of all the clusters in the region, and the
unusual cluster mass function, which has one massive cluster at least
10 times bigger than the rest in the midst of a swarm of what appear
to be more normal clusters. Whether other globular clusters or even
open clusters have had similar accretion histories is an intriguing
possibility.  Better observations at higher angular resolution are
necessary to check this model.

Acknowledgements:

Yu.E. appreciates partial support from the Russian Foundation for Basic
Research and the Council for Support of Scientific Schools. The research
of SSL was supported by the Danish Natural Science Research Council
through its Centre for Ground-Based Observational Astronomy. 
Many helpful suggestions by Tom Richtler are gratefully
acknowledged, as are useful comments by the referee.

\newpage

\newpage

\newpage

\newpage

\newpage

\newpage

\newpage
\end{document}